\input{aipcheck}

\documentclass[
    ,final            
    ,numberedheadings 
  ]
  {aipproc}

\newcommand{\be}{\begin{equation}}\newcommand{\ee}{\end{equation}}
\newcommand{\bear}{\begin{eqnarray}}\newcommand{\eear}{\end{eqnarray}} 
\newcommand{\ba}{\begin{array}}\newcommand{\ea}{\end{array}}

\layoutstyle{6x9}

\begin{document}

\title{Two Universal Extra Dimensions}

\classification{}
\keywords      {}

\author{Gustavo Burdman}{
  address={Instituto de F\'{i}sica, Universidade de S\~{a}o Paulo \\
R. do Mat\~{a}o, Travessa R, 187, Cidade Universitaria\\
S\~{a}o Paulo, SP 05580-900, Brazil}
}



\begin{abstract}
We review the main results of the 
study of the Standard model propagating in two universal extra
dimensions.  
Gauge bosons  give rise to heavy spin-1 and spin-0 particles.
The latter constitute the lightest Kaluza-Klein (KK) particle in the 
level (1,0). The level (1,1) can be s-channel produced. 
The main signals at the Tevatron and the LHC 
will be several $t\bar t$ resonances. 
\end{abstract}

\maketitle


\section{Introduction}
If compact universal extra dimensions (UED)
exist their compactification radius can be as large as few hundred
GeV$^{-1}$~\cite{ued1}. This is due to the restrictions imposed 
by KK parity. 

The degenerate KK spectrum is split by radiative 
corrections~\cite{uedrad}. However, these splittings are generically
small and make the phenomenology at hadron colliders rather 
non-trivial~\cite{fooled} due to the relatively small energy release in the 
decay chains. Alternatively, level 2 KK modes can be s-channel 
produced~\cite{res2}. 
Their couplings to zero modes are suppressed by volume (or loop) factors, 
but since they can decay through these couplings to light particles, 
they release enough energy. On the other hand, in the 5D case, the 
mass of the 2nd KK level is just $2/R$, twice the first level, up to radiative
corrections. Then, these excitations in principle can also decay to 
two level 1 KK modes, so that the clean decays of the  2nd KK level to 
two zero modes have to compete with these not-so-clean channels.

\section{Two Universal Extra Dimensions}
Considering the case with {\em two} Universal Extra Dimensions (TUED)
has several motivations. It is possible to protect 
the proton from decaying~\cite{pstable}, and requiring the cancellation
of chiral anomalies implies the existence of three families. 
However, here we concentrate on the phenomenological differences with
the case of one UED. 

There are two main aspects that differentiate TUED theories from 
the one UED case. First, the level-2 vector modes in the TUED 
offer better opportunities for
discovery  
since they have masses which are larger than the level-1 masses by a factor of
approximately $ \sqrt{2}$.  As a result, their production is possible
at smaller center-of-mass energies, and the decays of level-2 states
into pairs of level-1 states, which would lead to only soft leptons
and jets in the detector, are kinematically forbidden (as opposed to
the case of one UED where such decays are
typically allowed).  Then the level-2 states, 
characterized by KK numbers (1,1), have large branching fractions 
for decays into a pair of standard model particles giving rise to a
high $p_T$ signal.  
The second distinctive feature of TUED is that each vector mode 
is accompanied by a spin-0
particle in the adjoint representation of the corresponding gauge
group. These are the un-eaten Nambu-Goldstone bosons of the breaking 
of translation invariance. These two features will drive the phenomenology 
of the TUED to be quite distinctive. 

We compactify the TUED on a chiral square as defined in Refs.~\cite{chiralsq}.
It consists of a square, $0 \leq x_4, x_5 \leq \pi R$, where $x_4,x_5$
are the coordinates of the extra dimensions and $R$ is the
compactification ``radius''.  The compactification is obtained by
imposing the identification of two pairs of adjacent sides of the
square. 
The KK modes are characterized by two numbers, $j$ and $k$, 
integers with $j \geq 1$ and $k \geq
0$, or $j=k=0$.  The 4D fields $\Phi^{(j,k)}(x^\mu)$ are the KK modes
of the 6D field $\Phi(x,y)$.  They have masses due to the momentum along
$x^4, x^5$ given by
%
$M_{j,k} = \frac{1}{R} \sqrt{j^2 + k^2} ~,
$
so that the mass spectrum, in the limit where other contributions to
physical masses are neglected, starts with $M_{0,0} = 0$, $M_{1,0} =
1/R$, $M_{1,1} = \sqrt{2}/R$, $M_{2,0} = 2/R, \ldots$ 

The 6D gauge bosons decompose into a tower of 4D spin-1 fields, a
tower of 4D spin-0 fields that are eaten by the spin-1 states, plus 
a tower of spin-0 fields that {\em remain in the spectrum}. 
So there will be adjoint scalars: $G_H^{a(j,k)}$ corresponding 
to the 6D gluon field, and $W^{\pm(j,k)}_H$, 
$W^{3(j,k)}_H$ and $B^{(j,k)}$, corresponding the the electroweak 
6D fields. This is a crucial difference with the 5D case.
In what follows we summarize the results of Ref.~\cite{rested}, 
where a detailed discussion on the KK spectrum and couplings, production
cross sections and branching ratios can be found.

\subsection{Localized Operators}
The compactification on the chiral square implies the presence 
of three singularities. Operators localized on these
fixed points are generated at one loop by the bulk interactions, 
as well as by ultra-violet physics. We demand that the UV contributions
respect KK parity so as to keep the possibility of having a dark matter 
candidate. Then we have
\bear
\int_0^{\pi R} dx^4 \int_0^{\pi R} dx^5 \left\{ {\cal L}_{\rm bulk} +
\delta (x_4) \delta (\pi R - x_5) {\cal L}_2 \right.
\nonumber \\ \left.
\mbox{} + \left[ \delta (x_4) \delta (x_5) + \delta (\pi R - x_4) 
\delta (\pi R - x_5) \right]
{\cal L}_1 \right\} ~.
\label{l4d}
\eear 
The operators in ${\cal L}_1$ and ${\cal L}_2$ will give rise 
to KK number violating (but KK-parity conserving) contributions
to the KK masses and couplings. For instance, for gauge kinetic term
we have 
\be
- \frac{1}{4} \, \frac{C_{pG}}{\Lambda^2} 
G^{\mu\nu}G_{\mu\nu} 
- \frac{1}{2} \, \frac{C^\prime_{pG}}{\Lambda^2}
\left(G_{45}\right)^2 
\ee
with $p=1,2$, and similarly for fermion kinetic terms. 
Naive dimensional analysis can be used to estimate that the coefficients
of these operators are of order $C\sim \ell_6/\ell_4\simeq 8\pi$, where
$\ell_6$ and $\ell_4$ are loop factors in 6D and 4D respectively. 
On the other hand, the cutoff satisfies 
$\Lambda R\simeq (32/(\alpha_s N_c) \simeq 10$. This leads to the NDA estimate
of 
$\sim g^2/16\pi^2 
$
for the effective 4D couplings induced by the $C$'s.

Bulk interactions at one loop renormalize the localized operators below
$\Lambda$, generating large logarithms. 
The logarithmic enhancement, 
approximately $\ln(\Lambda R)\simeq 2.3$, is not particularly large.
However, in order to estimate the KK spectrum and the KK-number violating
couplings we will make use of these calculable contributions. 

For instance, we can compute the one-loop contributions to the KK mass 
spectrum.
One important finding is the fact that the lightest KK particle 
(LKP) of the first level (the level $(1,0)$), is the scalar adjoint
corresponding the the $U(1)_Y$ gauge boson, $B_H$. The rest of the spectrum 
for the first and second KK level can be seen in detail in Ref.~\cite{rested}.

\subsection{KK-Number-Violating Interactions}  

These interactions are generated by the localized operators.
For instance, the interactions of two zero-mode quarks with KK gluons 
is given by 
\be
g_s C^{qG}_{j,k} \left( \overline{q} \gamma^\mu T^a q\right)  
G_\mu^{(j,k)a} ~,
\label{coupling}
\ee 
where $g_s$ is the QCD coupling and, as we mentioned earlier, the 
coefficients such as $C^{qG}_{j,k}$ receive contributions from above $\Lambda$
as well as logarithmically enhanced contributions below the cutoff, 
which result
in $(g^2 N_c/16\pi^{2}) \ln (M^2_s/\mu^2)$. 
The typical size of these couplings are then this loop factor time log, times
a coefficient of order one, which can be estimated from the bulk one-loop 
contributions. 

The KK spinless adjoints interact with the zero-mode quarks only via
dimension-5 or higher operators:
\be
\frac{g_s \tilde{C}^{qG}_{j,k}}{M_{j,k}} \, 
\left(\overline{q} \gamma^\mu T^a q \right)\, D_\mu G_H^{(j,k)a}  
~,
\label{GH-coupling}
\ee
where $\tilde{C}^{qG}_{j,k}$ are real dimensionless parameters, and 
$D_\mu$ is the gauge covariant derivative.  
It is important
to notice that the vertex (\ref{GH-coupling}) is proportional to the
quark masses, as can be seen by integrating by parts and using the
fermion equations of motion.  As a result, the spinless adjoints decay
almost exclusively into top quarks.  This observation also implies
that the coupling for direct production of the spinless states is
negligible, being suppressed by the $u$ or $d$-quark masses.  However,
the spinless adjoints can be easily produced in the decays of KK
quarks or leptons. 

\section{Phenomenology of the 6D Standard Model}
The first KK level $(1,0)$ must be pair produced. On the other hand, 
the second KK level, $(1,1)$,  can be produced in the $s$ channel through 
KK-number violating interactions such as (\ref{coupling}) or 
(\ref{GH-coupling}). 
If we assume that the KK-number violating couplings are mostly 
radiatively induced, then the corresponding couplings of the 
electroweak level-2 KK modes  to zero-mode leptons is suppressed with respect
to the ones with zero-mode quarks: the electroweak level-2 KK modes are 
therefore mostly leptophobic. Although the level-2 KK gluon $G_\mu^{(1,1)}$ 
has a larger production cross section than the electroweak gauge bosons, 
its mass is also larger (is radiatively enhanced). Then, the number of events 
generated through level-2 KK gluons is comparable to the one originating 
from the cascade decays of electroweak level-2 KK gauge bosons. 

The fact that the dominant decay channels are hadronic, even for the 
electroweak KK modes, implies that the best hope for observation is in the 
$t\bar t$ channel. This is particularly true due to the fact that a 
significant fraction of the produced level-2 KK modes results in adjoint
scalars, which decay almost exclusively into $t\bar t$ pairs. The 
{\em five } resonances that decay to $t\bar t$, will form  {\em three}
distinguishable peaks: 
$G_H^{(1,1)}$ + $W_\mu^{3(1,1)}$, with  $\sim 1.10 \,{
M_{1,1}}$; 
$B_\mu^{(1,1)}$ + $W_H^{3(1,1)}$, with $\sim 0.97 \,
{M_{1,1}}$; and 
$B_H^{(1,1)}$, with $\sim 0.86 \,{M_{1,1}}$, 
where $M_{(1,1)}=\sqrt{2}/R$ is the level-2 un-corrected KK mass. 
In Ref.~\cite{rested} the reach of the Tevatron was studied.  
It was found that the Tevatron  can observe 
level-2 KK modes in the $t\bar t$ channel up to about $700~$GeV.
The ultimate Tevatron reach for the compactification scale 
is found to e about $R^{-1} \sim 500$~TeV. The production cross 
sections at the LHC were estimated using $q\bar q$ initial states only. 
It was found that the LHC should have a considerably larger reach on $R^{-1}$.
However, more detailed work is needed to include all contributions as 
well as a serious study of the backgrounds. 

Finally, a comment about the viability of the SM in UED and TUED. 
It was recently pointed in Ref.~\cite{lhued} that the SM in 5D UED has a 
``little hierarchy'' problem since the Higgs mass still receives 
quadratically divergent contributions. Solving this problem 
in the 5D scenario by assuming the Higgs is a pseudo-Nambu-Goldstone boson, 
results in the  cutoff dependence becoming linear. The phenomenology might 
change with respect to the SM in 5D depending on the mechanism used to solve 
the little hierarchy. However, in the 6D case of TUED, the cutoff dependence 
remains quadratic. Then more work is needed to address the stabilization 
of the Higgs mass in a theory with two inversal extra dimensions.








\bibliographystyle{aipproc}   


\IfFileExists{\jobname.bbl}{}
 {\typeout{}
  \typeout{******************************************}
  \typeout{** Please run "bibtex \jobname" to optain}
  \typeout{** the bibliography and then re-run LaTeX}
  \typeout{** twice to fix the references!}
  \typeout{******************************************}
  \typeout{}
 }


\end{document}